\documentstyle[11pt,newpasp,twoside,epsf,psfig]{article}
\def\pks{PKS\,1343$-$601} 
\def\kms{{\,km\,s$^{-1}$}} 

\def\edcomment#1{\iffalse\marginpar{\raggedright\sl#1\/}\else\relax\fi}
\reversemarginpar

\begin{document}
\title{A 2 x 2 Degree I-band Survey around PKS\,1343$-$601}
\author{Ren\'ee C. Kraan-Korteweg} 
\affil{Depto.\ de Astronom\' \i a, Universidad de Guanajuato,
Apdo.~Postal 144, Guanajuato, GTO 36000, M\'exico}
\author{Manuel Ochoa}
\affil{Depto.\ de Astronom\' \i a, Universidad de Guanajuato,
Apdo.~Postal 144, Guanajuato, GTO 36000, M\'exico \\
Instituto de Astronom\' \i a, UNAM, Apdo. Postal 70-264, M\'exico 
D.F. 04510, M\'exico}
\author{Patrick. A. Woudt}
\affil{Dept. of Astronomy, Univ. of Cape Town, Private Bag, 
Rondebosch 7700, South Africa}
\author{Heinz Andernach}
\affil{Depto.\ de Astronom\' \i a, Universidad de Guanajuato,
Apdo.~Postal 144, Guanajuato, GTO 36000, M\'exico}

\begin{abstract}
Motivated by the possibility that the highly obscured ($A_{\rm B} =
12\fm$) radio galaxy PKS\,1343$-$601 at $(\ell,b,cz) = (309\fdg7,
+1\fdg8, 3872$\,km\,s$^{-1}$) might constitute the center of a heavily
obscured cluster in the Great Attractor region, we have imaged about
$2\deg \times 2\deg$ of the core of this prospective cluster in the
$I$-band using the WFI at the ESO 2.2\,m telescope at La Silla. We
were able to identify 49 galaxies and 6 uncertain galaxy
candidates. Although their distribution does not resemble a centrally
condensed, massive cluster, its appearance -- severely influenced by
the strong dust gradient across our surveyed region -- is entirely
consistent with a cluster.
\end{abstract}

\section{Introduction}

The Great Attractor (GA), a large extended mass overdensity in the
nearby Universe, was discovered from the infall pattern of elliptical
galaxies (Dressler et al. 1987). ~Kolatt et al.~(1995) determined its
center and found it to ly exactly behind the southern Milky Way at
($\ell, b, cz) = (320\deg, 0\deg, 4000$\kms). Because of the
increasing dust absorption at lower Galactic latitudes it has remained
difficult to assess whether this mass density is being traced by the
galaxy distribution.

The deep optical galaxy search of partially obscured galaxies in the
GA region (Woudt \& Kraan-Korteweg 2001) revealed the Norma cluster
(ACO 3627) at ($\ell,b) = (325.3\deg,-7.2\deg$) to be a region of very
high galaxy density in the GA region. Follow-up redshift observations
found the Norma cluster to be comparable in size, richness and mass to
the Coma cluster, albeit nearer by a factor of 1.4 (Kraan-Korteweg et
al.~1996; Woudt et al.~these proceedings). It is therefore the most
likely candidate to constitute the previously unrecognized bottom of
the potential well of the Great Attractor (GA) overdensity.

Is Norma the only massive cluster behind the Milky Way in the GA
region or might other clusters form part of a much broader and hence
more massive core of the GA? The identification of possible further
dynamically important mass contributors to the GA at lower latitudes
is problematic, however, because of the high extinction in the
optical, star-crowding in the near-infrared surveys DENIS and
2MASS, and even in X-rays -- despite the fact that dust extinction
and stellar confusion are unimportant -- because of the photoelectric
aborption at high Galactic HI column densities.

\subsection{The \pks\ Galaxy}

PKS\,1343$-$601 is the 2nd strongest radio continuum source in the
southern sky ($f_{\rm 20cm} = 79$\,Jy; Mc Adam 1991 and references
therein) and lies at very low latitudes (${\ell},b$) = ($309\fdg7,
+1\fdg8$). Woudt (1998) and Kraan-Korteweg \& Woudt (1999) did suspect
that this radio galaxy might constitute the central galaxy of a
cluster: this galaxy lies behind an obscuration layer of about
12$^{\rm m}$ extinction in the B-band, as estimated from the
DIRBE/IRAS extinction maps (Schlegel, Finkbeiner, \& Davis 1998) and
is not visible in the optical. With a diameter of 28 arcsec in the
Gunn-z filter and, based on the H$\alpha$ emission line, a recession
velocity of $cz = 3872$\kms\ (West \& Tarenghi 1989), it must be a
giant elliptical galaxy (a diameter of about 4 arcmin if corrected for
extinction effects using Cameron's (1990) extinction laws). Such
galaxies are not generally isolated but found at the cores of clusters.

If \pks\ marks the dynamical center of a cluster, then the Abell
radius, defined as 1$\farcm$7/$z$ where $z$ is the redshift,
corresponds to $R_{\rm A} = 2\fdg2 = 3\,h_{50}^{-1}$\,Mpc) on the sky
at the redshift-distance of \pks.  In the optical, only a handful of
highly obscured galaxies (at the highest latitudes and lowest
extinction levels) could be identified within this radius and even
2MASS reveals only 9 galaxy candidates within this radius.

As rich clusters generally are strong X-ray emitters, we searched for
evidence of such emission. \pks\ has not been detected with ROSAT but
the soft X-ray emission would be strongly absorbed by the Galactic HI
at that position. It has, however, been detected with ASCA (Tashiro et
al.~1998) showing slightly extended diffuse hard X-ray emission at the
position of \pks. The flux of kT = 3.9~keV is quite high for a single
galaxy, hence could be indicative of emission from a cluster. However,
recent higher resolution X-ray observations with XMM do not support
this supposition (see Schr\"oder et al., these proceedings). The
extended X-ray emission is thus probably due to the Inverse Compton
process in the radio lobes, as already suggested by Ebeling, Mullis,
\& Tully in 2002.

Interestingly enough, the ZOA Parkes Multi Beam H\,I survey does
uncover a significant excess of galaxies at this position in velocity
space (Kraan-Korteweg et al.~2004; Henning, Kraan-Korteweg, \&
Staveley-Smith, these proceedings). However, no ``finger of God'' is
obvious, the characteristic signature of a cluster in redshift space.
Then again, HI is not a good tracer of high-density regions such as
cluster cores, since spiral galaxies generally avoid the cores of
clusters or are HI depleted (Bravo-Alfaro et al.~2000, Vollmer et
al.~2001).

The HI-velocities plus the few known optical velocities within the
Abell radius and its immediate surroundings provide some dynamical
support for the existence of this cluster. Not only do we find a
significant peak in the velocity histogram at the velocity of \pks,
but the velocities as a function of distance from the cluster center
lie within a narrow range of the central radio source and are well 
separated in velocity space from field galaxies (see Fig.~1).

\begin{figure}
\hfil \psfig {figure=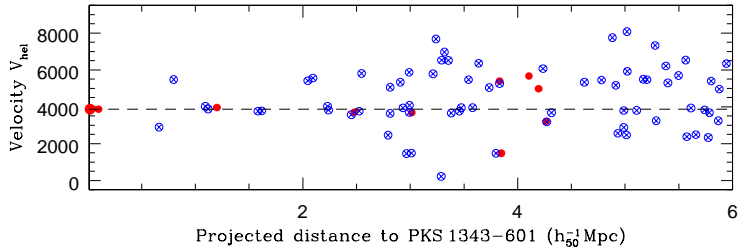,width=10cm} \hfil
\caption{The velocity distribution as a function of distance from
\pks. The crossed circles are galaxies detected in the Parkes ZOA
Multibeam Survey (Henning et al., in prep.), the filled circles from
optical velocity data as given in LEDA.}
\label{R_vel}
\end{figure}

To verify whether \pks\ indeed marks the center of an obscured
cluster, we have imaged the core of this prospective cluster in the
$I$-band in which extinction effects are less severe ($A_{\rm I} =
0.45 A_{\rm B}$) using the Wide Field Imager WFI of the ESO/MPG 2.2-m
telescope at la Silla.

\section{The 2 x 2 Degree $I$-band Survey}

A total of 16 WFI fields, each $34\arcmin \times 33\arcmin$, covering
a total area of about $2\deg \times 2\deg$ were observed in May
1999. The surveyed area is outlined in Figs.~3 and 4 together with the
Abell radius of the prospective cluster. The field centered on \pks\
has a slightly higher exposure time (1500s compared to 600s) and
consists of 5 dithered exposures to improve the spatial resolution.

After standard reduction with $IRAF$, all the images were inspected
visually. The small images of the strongly obscured galaxies and the
heavy star-crowding (partly covering the galaxies) make an automatic
detection algorithm impossible. A constant changing of the intensity
and contrast levels while viewing a field brings out the more
extended low surface brightness borders of galaxies and also emphasizes
the difference in the light distribution between stars and
galaxies. This procedure makes the identification of such obscured
objects more feasible.

In this way, 49 galaxies were identified, 25 of them probably of
elliptical morphology, in addition to 6 uncertain galaxy candidates.
A sample of galaxy images, from the brightest to the faintest
galaxies, is displayed in Fig.~2.

\begin{figure}[!t]
\hfil \psfig {figure=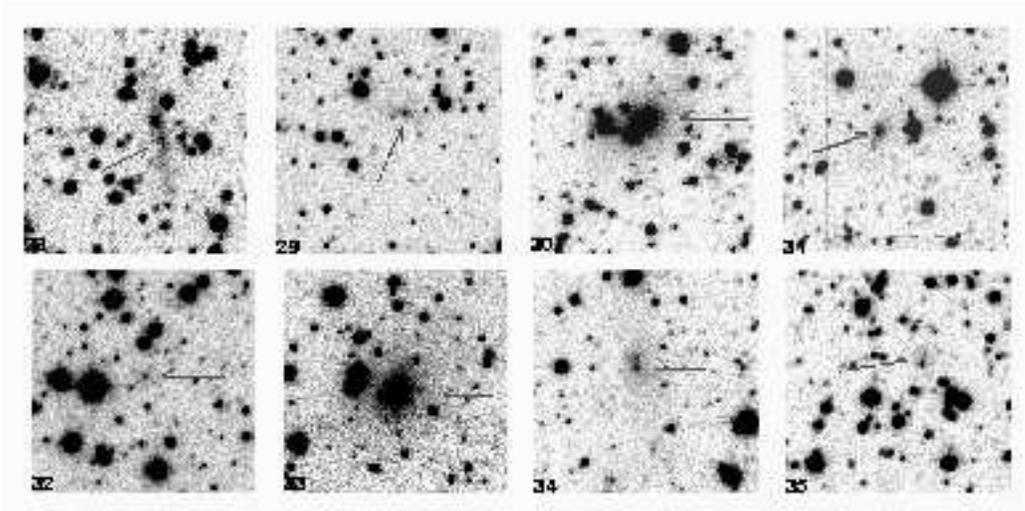} \hfil
\caption{An image gallery of 8 of the 49 certain galaxy
candidates identified around \pks\ on the $I$-band images. 
Each image is about $36\arcsec\times 36\arcsec$.}
\label{image}
\end{figure}

Galaxy 33 is the giant radio galaxy PKS\,1343$-$601 ($A_{\rm I} =
5\fm5$).  The spiral galaxy (\#\,28) is by far the largest galaxy
identified in this survey but it is located in an area of relatively
low extinction ($A_{\rm I} = 2\fm2$). Galaxies 29, 30, 32, 33 and 34
are from the central field and although some of them look like mere
smudges (e.g. \#\,32) all of them are independently confirmed by a NIR
survey ($J, H, K '$) of the central $36\arcmin \times 36\arcmin$
region of this possible cluster using the Japanese/South African
Infrared Survey Facility (Nagayama et al.~2004; Nagayama at al., these
proceedings).  In fact these recent observations reveal the very low
surface brightness galaxy \#\,34 at $A_{\rm I} = 5\fm5$ to actually be
a very extended edge-on spiral galaxy.

\section{The Detected Galaxy Distribution}

The distribution in Galactic coordinates of the unveiled galaxies is
shown in Fig.~3. The outlined square region indicates the area imaged
with the WFI in the $I$-band (16 fields of $34\arcmin \times
33\arcmin$ each) around \pks\ (large dot), and the circle the Abell
radius of $R_A = 2\fdg2 = 3\,h_{50}^{-1}$\,Mpc. The filled circles
mark the 25 elliptical galaxies and the crossed circles the 24 spiral
galaxies.  Note that the morphology is very uncertain due to the heavy
obscuration. The open circles show another 6 uncertain candidates.

\begin{figure}
\hfil \psfig {figure=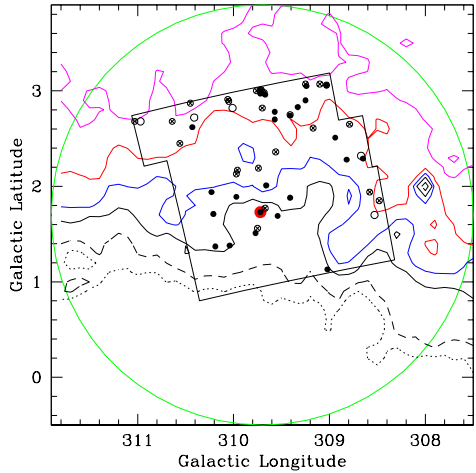,width=6cm} \hfil
\caption{The distribution in Galactic coordinates of the identified
galaxies. The surveyed area is marked as the tilted square. The circle
corresponds to the Abell radius. Filled, crossed and open circles mark
elliptical, spiral and uncertain galaxies. The contours indicate
extinction levels in the $I$-band: $A_I=2\fm0, 3\fm0, 4\fm0, 5\fm0$
(solid), $7\fm5$ (long dash) and $10\fm0$ (short dash) following
Schlegel et al.~1998.}
\label{dist}
\end{figure}

The distribution shows that is was possible to identify galaxies to
extinction levels of $A_{\rm I} \la 5\fm0$, i.e., to similar
extinction levels within this photometric passband compared to our
deep optical galaxy searches in the $B$-band where we identified
galaxies down to $A_{\rm B} \la 5\fm0$ (Kraan-Korteweg 2000; Woudt \&
Kraan-Korteweg 2001). Ellipticals are found mainly at higher
extinction levels while at lower extinction levels spirals
predominate. As the cores of ellipticals and bulges of spirals contain
mainly the red old star population this trend was to be expected.  It
is still interesting to note that the density of ellipticals seems to
show a concentration around \pks, even though the distribution does
not match our expectations of a normal, centrally condensed, rich
cluster. Its appearance is, however, strongly modulated by the
extinction gradient. To assess whether the unveiled distribution
really is consistent with a rich cluster hidden behind an increasingly
thickening exinction layer, we simulated how a rich cluster would
appear at the position of \pks.

\section{Does the Galaxy Distribution Indicate a Galaxy Cluster?}

For the simulation, we have used the deep Coma cluster catalog of 6724
galaxies by Godwin, Metcalfe, \& Peach (1983). First we move the rich
Coma cluster at the radial velocity of \pks. This results in an
extension of the cluster size and galaxy diameters by a factor of
$f=1.77$ and an increase in brightness of $\Delta m = 1\fm24$. We then
transform the $B$-band magnitudes to the $I$-band with a mean color
term of $(B-I) = 2\fm0$ which is representative of early-type galaxies
as well as bulges of spiral galaxies. Although it is difficult to
assess the magnitude limit obtained with the here used non-standard
narrow $I$-band filter ($\Delta \lambda = 275$\,\AA\ centered on
$\lambda = 9148\,$\AA), we estimate to be able to find galaxies to a
standard $I$-band magnitude of $I_{\rm lim} \approx 17\fm5$. Applying
this cut-off would leave us with 1578 galaxies in our surveyed cluster
area. The left panel in Fig.~4 shows the resulting distribution, where
the symbols are proportional to brightness.

\begin{figure}
\hfil \psfig {figure=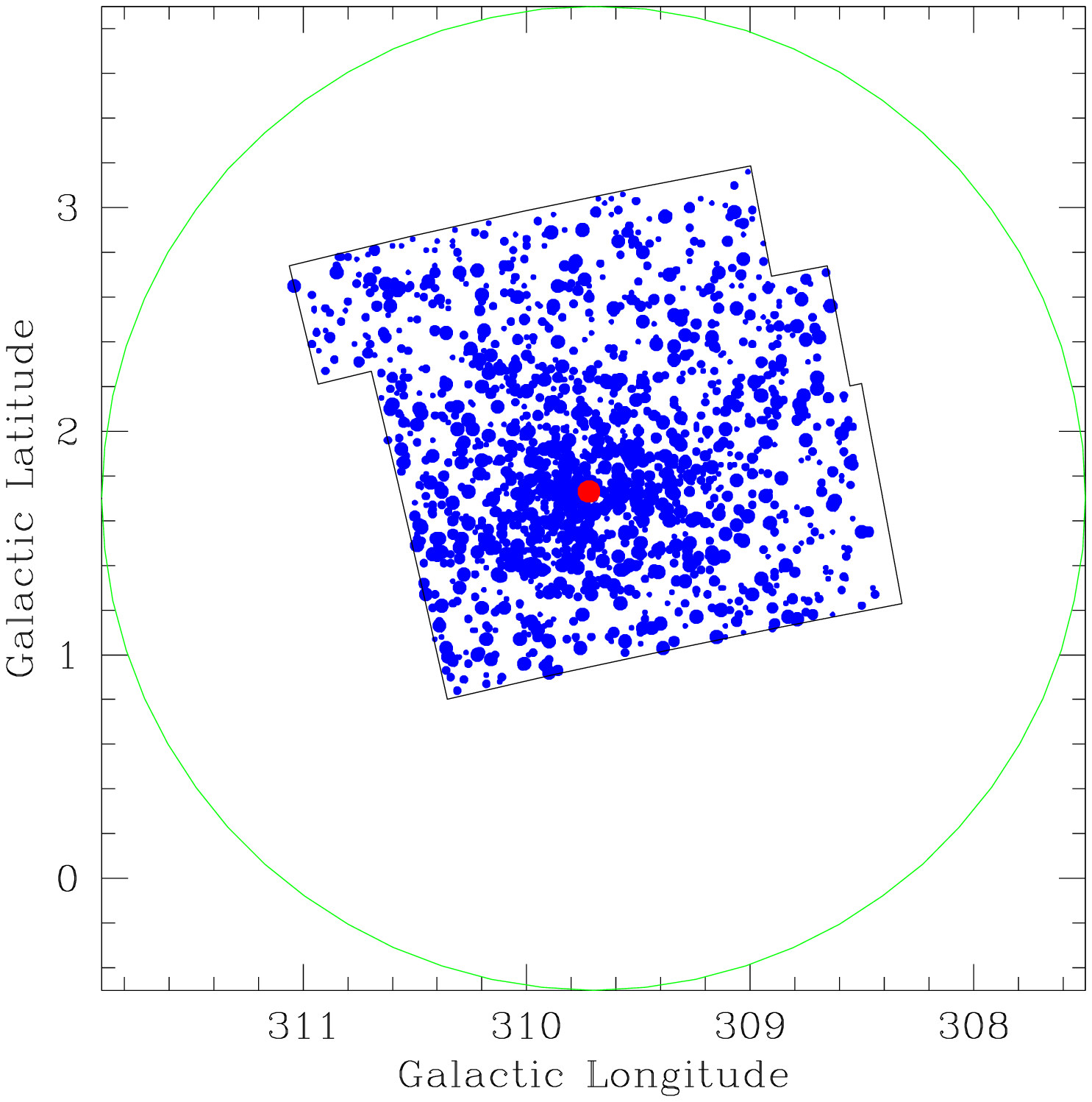,width=6cm} 
\hfil \psfig {figure=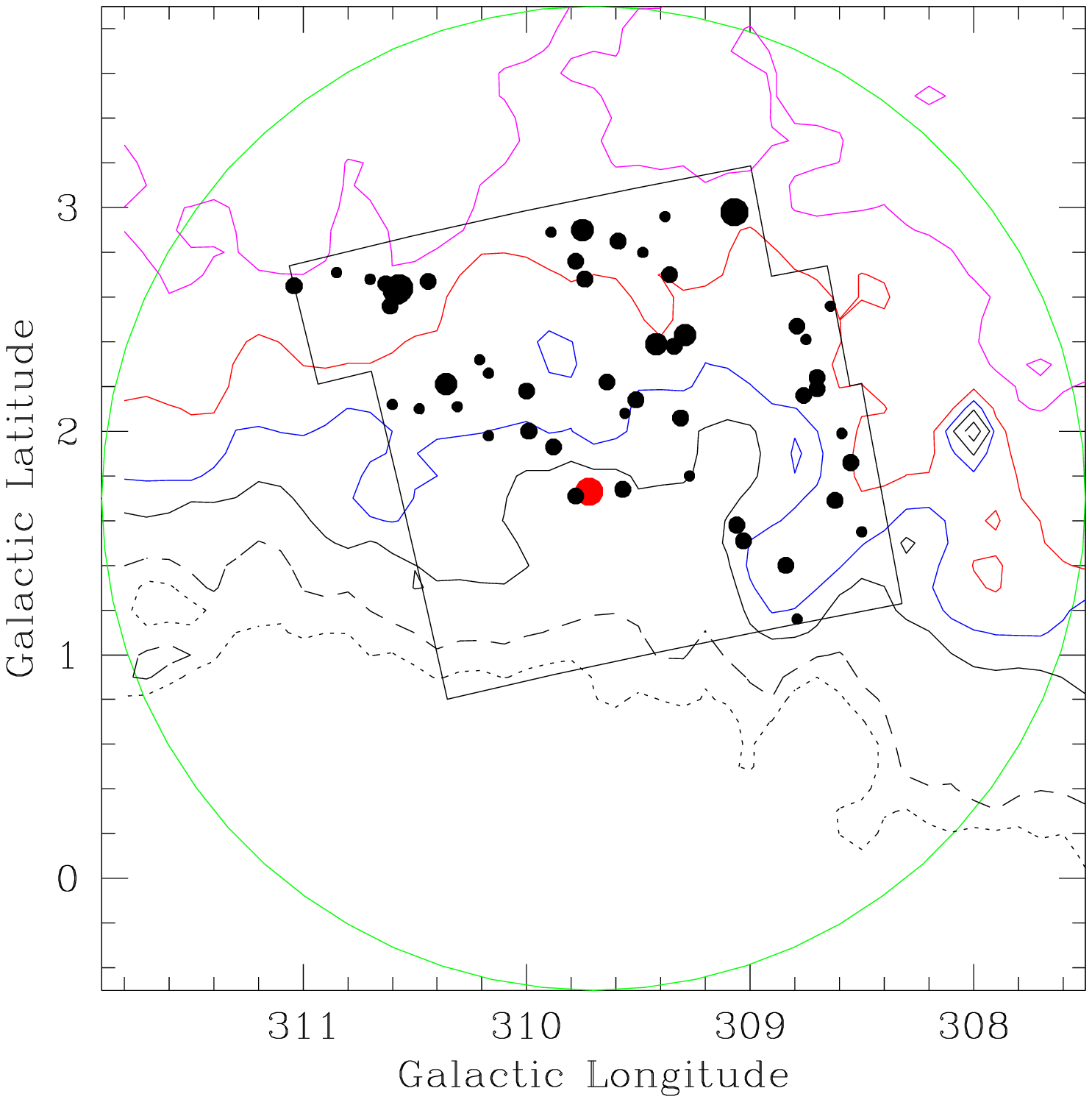,width=6cm} 
\caption{Left panel: Simulation of the distribution of Coma cluster
galaxies if it would be positioned at the redshift space of \pks, and
would have been observed under the same conditions with the with the
Wide Field $I$-band imager at La Silla. Right panel: Remaining
galaxies if the cluster galaxies are subjected to the same foreground
extinction (extinction contours as in Fig.~3).}
\end{figure}

However, we still have to take the effect of dust absorption into
account.  Even in the $I$-band, the minimum absorption is $A_{\rm I} 
\approx 2^{\rm m}$ increasing to a maximum of $A_{\rm I} \approx 10^{\rm
m}$. Absorbing each galaxy individually according to the DIRBE
extinction value at their respective positions (see contour levels in
Fig.~3) using the inverse Cameron laws (1990), leaves only 51
identifiable galaxies. Their distribution is shown in the right panel
of Fig.~4.

This number is entirely consistent with the number of galaxies
identified in our real search (49 certain and 6 uncertain galaxy
candidates). The resulting distribution furthermore is a near replica
of the actual detections (compare right panel of Fig.~4 with Fig.~3):
galaxies are found down to the same extinction level of about $A_I \la
5{\rm m}$. Only Coma's two cD galaxies peak through slightly thicker
extinction layers -- similar to \pks\ -- and the galaxy numbers within
the various extinction intervals are quite similar.

\section{Discussion}

The simulation thus seems to suggest that the uncovered galaxy
distribution in our $I$-band survey around \pks\ is consistent with
being due to a galaxy cluster at the position and distance of
\pks. Care should, however, be taken when interpreting the galaxy
simulations. Small changes in, e.g. the estimated $I$-band magnitude
limit, the actual distance, the Cameron (1990) extinction corrections
(which are type-dependent and determined for the $B$-band), assumption
of a fixed $B-I$ color term, etc., might alter the simulated galaxy
distribution significantly. Furthermore, the fact that no X-ray
emission typical of a massive cluster (Ebeling et al. 2002, Ebeling et
al., these proceedings, Schr\"oder et al., these proceedings) argues
against a very rich cluster. The most likely interpretation therefore
is, that there exists a cluster around \pks, but that it rather is an
intermediate-mass cluster like Hydra or Centauraus, which is in
agreement with the independent analysis by Nagayama et al.~2004,
Nagayama et al., these proceedings.

\acknowledgements This research used the Lyon-Meudon Extragalactic
Database (LEDA), supplied by the LEDA team at the Centre de Recherche
Astronomique de Lyon, Obs.~de Lyon. RCKK, MO and HA thank CONACyT for
their support (research grants 27602E and 40094F). PAW acknowledges
the National Research Foundation for financial support.

\end{document}